\documentclass[useAMS,usenatbib,twocolumn]{mn2e}

\usepackage{amsmath,amssymb}
\usepackage{subfigure}
\usepackage{graphicx}
\usepackage{natbib}
\usepackage{url}

\newcommand{\spose}[1]{\hbox to 0pt{#1\hss}}
\newcommand{\lta}{\mathrel{\spose{\lower 3pt\hbox{$\mathchar"218$}}
     \raise 2.0pt\hbox{$\mathchar"13C$}}}
\newcommand{\gta}{\mathrel{\spose{\lower 3pt\hbox{$\mathchar"218$}}
     \raise 2.0pt\hbox{$\mathchar"13E$}}}

\newcommand{\erf}{\mathop{\rm erf}\nolimits}

\newcommand{\aj}{AJ}         
\newcommand{\apj}{ApJ}       
\newcommand{\mnras}{MNRAS}   

\title[Direct Simulation Monte Carlo]{Direct Simulation Monte Carlo for
  astrophysical flows: I. Motivation and methodology}

\author[M.~D.~Weinberg]{Martin D. Weinberg\thanks{E-mail:
    weinberg@astro.umass.edu}\footnotemark[1] \\
  Department of Astronomy\\ University of Massachusetts, Amherst
  MA 01003-9305, USA }

\begin{document}

\label{firstpage}

\pagerange{\pageref{firstpage}--\pageref{lastpage}} \pubyear{2013}

\maketitle

\begin{abstract}
  We describe a hybrid Direct Simulation Monte Carlo (DSMC) code for
  simultaneously solving the \emph{collisional} Boltzmann equation for
  gas and the \emph{collisionless} Boltzmann equation for stars and
  dark matter for problems important to galaxy evolution.  This
  project is motivated by the need to understand the controlling
  dynamics at interfaces between gases of widely differing densities
  and temperature, i.e. multiphase media.  While more expensive than
  hydrodynamics, the kinetic approach does not suffer from
  discontinuities and it applies when the continuum limit does not,
  such as in the collapse of galaxy clusters and at the interface
  between coronal halo gas and a thin neutral gas layer.  Finally, the
  momentum flux is carried, self-consistently, by particles and this
  approach explicitly resolves and thereby `captures' shocks.
  
  The DSMC method splits the solution into two pieces: 1) the
  evolution of the phase-space flow \emph{without} collisions; and 2)
  the evolution governed the collision term alone \emph{without}
  phase-space flow.  This splitting approach makes DSMC an ideal match
  to existing particle-based n-body codes.  If the mean free path
  becomes very small compared to any scale of interest, the method
  abandons simulated particle collisions and simply adopts the relaxed
  solution in each interaction cell consistent with the overall energy
  and momentum fluxes.  This is functionally equivalent to solving the
  Navier-Stokes equations on a mesh.  Our implementation is tested
  using the Sod shock tube problem and the non-linear development of
  an Kelvin-Helmholtz unstable shear layer.
\end{abstract}


\section{Introduction}
\label{sec:intro}

\subsection{Motivation}
\label{sec:motivation}

The burgeoning volume of multiwavelength galaxy observations reveals
an interactive complexity of patterns that couple merging,
environmental interaction, enhanced epochs of star-formation, and gas
accretion histories \citep[e.g.][]{Mo.etal:2010}.  Each mass
component---dark matter, stellar and compact objects, atomic and
molecular gas---is sensitive to different although overlapping ranges
of length and time scales.  Key to piecing together these complex
interactions is an understanding of the mutual dynamical evolution of
each component.  For gas in particular, winds, accretion flows and
other collisionally-induced structures often produce phase interfaces
at shocks.  Not only do the density enhancements at these shocks
result in radiative cooling and instabilities, these enhancements
provide gravitational \emph{handles} for momentum transfer and
torques.

More generally, galaxies are particularly sensitive to the conditions
in \emph{transitional} regimes.  These transitions occur in density
(e.g. between galaxy components), gas temperature and gravitational
acceleration.  For example, on the largest scales, the standard model
predicts that gravitational potential of a galaxy is dominated by dark
matter.  The baryons in the halo are rarefied with a temperature
characteristic of the escape velocity from the halo, approximately
$10^6$ degrees K.  Moving inwards, there is a transition between the
baryon-dominated stellar and gaseous galaxy and the collisionless dark
halo.  In this region, the gravitational dynamics of both components
may conspire to produce structure.  The increased gas density shortens
the atomic and molecular cooling time, producing a dense cold gas
layer.  This implies that there must be an interface of coexistence
between the discrete rarefied and dense phases.  Finally, each of
these gas phases and the gravitationally dominant collisionless
components has a characteristic length and time scale that produces a
range of accelerations.
 
These transitions regions provide tests of the standard and
alternative galaxy formation hypotheses.  In particular, numerical
simulations based on the standard $\Lambda$CDM scenario have made a
variety of predictions on small galactic scales with both successes
and failures.  However, first-principle physical simulations are often
infeasible at the scales where many of the failures occur, so at this
point, the failures may either intrinsic or methodological.
Alternately, some propose that the root cause of the failures may be
the nature of gravity itself and that \emph{modified Newtonian
  dynamics} (MOND) produces a better prediction of observed rotations
curves than standard $\Lambda$CDM.  Since the outer galaxy will be
dominated by dark matter in the standard scenario or by non-Newtonian
forces in MOND, the response in the outer galaxy to the dynamics may
provide a sensitive test for these hypotheses.  In addition, the
temporal differences between the collisionless and collisional
responses of media within each galaxy component provides clues to the
history of a galaxy's evolution that depend on the underlying
cosmogony.  For example, the gas responds over several hundred million
years to a disturbance in the stellar and dark-matter response to an
external perturbation taking place over gigayears.  The observation of
both responses \emph{together} can provide key diagnostics to the
evolutionary history and possibly the underlying physics.

Unfortunately, these transition regions are difficult to simulate
reliably.  Therefore, the most productive use of the combined
DSMC--n-body hybrid code will be the investigation of interface
dynamics and energetics on small and intermediate scales.  An
experimental version of our code uses the full
CHIANTI\footnote{CHIANTI is a collaborative project involving George
  Mason University, the University of Michigan (USA) and the
  University of Cambridge (UK).} atomic database
\citep{Dere.etal:1997,Landi.etal:2011} for collisional cross sections
and recombination (bound-bound and bound-free) and standard plasma
transitions (free-free) for the ionized regime.  The current version
does not treat the electrons as a separate kinetic species, but rather
assumes that they follow the ions.  This restriction will be relaxed
in a future version of this code and this will allow a fully
self-consistent treatment of electron conduction and allow the
dynamical influence of fixed magnetic fields to be investigated.  DSMC
could be incorporated into a full-fledged PIC plasma code
\citep[e.g.][]{Serikov.etal:1999} but this is well beyond the scope of
our current implementation.

Our main purpose for this and the companion paper \citep[hereafter
Paper 2]{Weinberg:13b} is a demonstration of the value of the kinetic
approach for astrophysical flows by applying it to some classic
scenarios.  To facilitate this comparison, we use the standard,
simplified local thermodynamic equilibrium (LTE) scheme
\citep[e.g.][]{Black:81} rather than the full-fledged self-consistent
cross-section based approach.  We will begin, in the next section,
with a very brief overview of the different regimes for gas dynamics
from rarefied to dense. This will motivate the need for an
understanding of gas in the transitional regime.  Section
\ref{sec:method} describes the numerical approach in two parts:
\S\ref{sec:DSMC} introduces the DSMC algorithm and a hybrid variant
that exploits the near-equilibrium solution when the mean-free path is
very short, and \S\ref{sec:implement} briefly reviews the n-body code,
describes the implementation of the DSMC algorithm in parallel using
MPI \citep{Gropp.etal:1999a,Gabriel.etal:2004} and presents
diagnostics necessary for parameter tuning.  We present two code tests
in \S\ref{sec:tests}: the standard one-dimensional shock tube
(\S\ref{sec:shock}) and the recovery of the Kelvin-Helmholtz
instability (\S\ref{sec:KH}).  We conclude in \S\ref{sec:summary} with
a discussion and summary.

\section{Gas simulation}
\label{sec:gas_sim}

Gas dynamics in galaxies is most often simulated through numerical
solutions of the Navier-Stokes equations:
\begin{equation}
  \rho \left(\frac{\partial \mathbf{v}}{\partial t} + \mathbf{v}
    \cdot \nabla \mathbf{v} \right) = -\nabla P + \nabla
  \cdot\mathbb{T} + \mathbf{F},
\label{eq:nse}
\end{equation}
where $\mathbf{v}$ is the flow velocity, $\rho$ is the fluid density,
$P$ is the pressure, $\mathbb{T}$ is the stress tensor, and
$\mathbf{F}$ represents body forces (per unit volume) acting on the
fluid.  In essence, this equation is an application of Newton's second
law to a continuum and is most often derived from this point of view.
However, to truly begin with Newton's Laws requires the application of
kinetic theory with the Boltzmann equation as a starting point
\citep[e.g. see][]{Fritz:2001}.

The Navier-Stokes equations (eq. \ref{eq:nse}) are rife with shock
discontinuities in general and are notoriously difficult to solve.
Physically, the width of the shock interfaces are of order the
mean-free path, and therefore are formally inconsistent with the
continuum approximation.  On the other hand, these shock
discontinuities are responsible for driving important astrophysical
phenomena on many scales (thermal heating, chemistry and radiation,
turbulence, to name a few) and must be treated carefully.  Because of
this, much of the difficult work in computational fluid dynamics
concerns the approximations at interfaces.  For example, grid and
finite element-schemes use shock-capturing methods to stabilize the
solution in presence of discontinuities.  These methods often
introduce numerical dissipation to achieve stability and add some
\emph{width} to the discontinuity prevent numerically-induced
oscillations.  The SPH method requires the introduction of artificial
dissipation terms that enable the conservation energy and momentum at
the otherwise unresolved shock discontinuity.  Therefore, the
effective width of the shock will most often not correspond to the
intrinsic width which is of order the mean-free path.  Although
artificial viscosity allows the discontinuity to be resolved, the
algorithm may also introduce unphysical dynamics.  For example,
\cite{Agertz.etal:07} illustrated the appearance of a smooth ordered
layer of particles near discontinuities that inhibits Kelvin-Helmholtz
and Rayleigh-Taylor instabilities near density gradients.  A number of
fixes have been recently introduced to help address this problem
\citep{Read.etal:10, Hopkins:13}.  In the case of discontinuous
Galerkin methods, one may use order reduction or limiting to prevent
spurious oscillations near discontinuities.  Spectral methods, which
project the fluid equations onto basis functions (e.g. spherical
harmonics, Chebyshev polynomials), yield high-accuracy solutions but
one must be \emph{even more} careful at interfaces.  Current state of
the art in spectral methods is to use high-accuracy shock-fitting
algorithms.  In summary, when the interactions in the shock interface
are critical to the resulting energetics and observational
diagnostics, artificial viscosity and shock-capturing techniques which
rely on approximating the discontinuous nature of the shock will miss
important physics.

A more general formulation of gas dynamics follows from the
collisional Boltzmann equation,
\begin{equation}
  \frac{\partial f}{\partial t} + \mathbf{v}\cdot{\mathbf{\nabla}_{\bf
      x} f} + \mathbf{F}\cdot{\mathbf{\nabla}_{\bf v} f} = 
  Q(f, f)
  \label{eq:be}
\end{equation}
which describes the change to the phase-space density, $f$, induced by
the collisions between particles, $Q(f,f)$.  Since $f$ has a
six-dimensional domain and the fields in the Navier-Stokes equation
have a three-dimensional domain, equation (\ref{eq:be}) appears harder
and certainly more time consuming to solve than equation
(\ref{eq:nse}).  However, the high level of algorithmic complexity in
computational fluid dynamics (CFD) follows from the mathematical
requirements that arise from the maintenance of the continuum limit.
On the other hand, the left-hand side of equation (\ref{eq:be}) is the
collisionless part of the Boltzmann that may be readily solved using
n-body techniques while the left-hand describes the particle
collisions.  Unlike the equations of hydrodynamics, the collisional
Boltzmann equation has \emph{no problem} with transition regimes or
discontinuities since the information about the collisions are carried
ballistically rather than by constitutive relationships.  Said another
way, the continuum limit will often fail by definition at transition
regimes while the kinetic approach represents the physical nature of
the transition and cannot fail. The price for this generalization is
performance: the kinetic approach requires smaller timescales and
length scales. Solutions of the collisional Boltzmann equation are
often an order of magnitude slower or more than hydrodynamic solutions
for the same problem.

\begin{figure}
  \centering
  \includegraphics[width=0.5\textwidth]{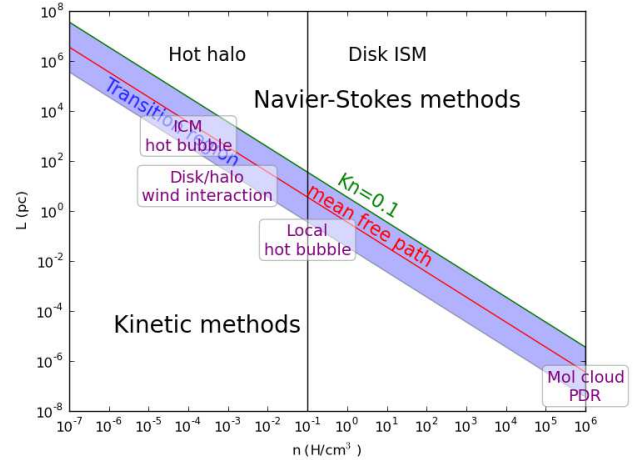}
  \caption{Gas regimes as a function of atomic density in atoms/cc and
    characteristic scale in parsecs.  The $Kn=1$ line is shown in red.
    The shaded blue region is the transition region.  For $Kn\gta10$,
    the flow is strongly in the kinetic regime, where collisions and
    excitation may be important overall but rare.  For $Kn\lta0.04$,
    the continuum limit, e.g. the Navier-Stokes equation, is
    appropriate.  A number of classic astronomical interface regimes
    are shown in labeled boxes.}
  \label{fig:regimes}
\end{figure}

Given the difficulties in solving equation (\ref{eq:nse}) and its
inappropriateness for true transitional regimes, it is worth exploring
alternatives.  Direct solution of the collisional Boltzmann equation
(\ref{eq:be}) using a kinetic simulation method is attractive for a
number of reasons.  The simplest kinetic method, molecular dynamics,
solves the full n-body system including the collisions directly, at
often great expense.  On the other hand, it is straightforward to
incorporate any number of species and specific physical couplings that
would be difficult in CFD, and generalization to multiple species and
interactions is straightforward.  Several alternatives to pure
molecular dynamics exist; each achieves computational efficiency by
approximating some aspect of equation (\ref{eq:be}).  For example,
lattice Boltzmann methods (LBM) discretize and solve equation
(\ref{eq:be}) on a grid and naturally yield the Navier-Stokes equation
in the small mean-free path limit.  This is a form of mesoscopic
solution designed to produce the correct solution on scales larger
than the particle scale but smaller than the continuum macroscopic
scale.  The final example of a mesoscopic solution is Monte Carlo
solution of Boltzmann equation.  This approach exploits the classical
indeterminacy of the collisional trajectories on the mean-free-path
scale.  The effects of the collisions are incorporated with a
Monte-Carlo procedure that reproduces the per-particle cross sections
that affect the flow at intermediate length scales.  This method of
solution may be one of the few practical approaches available to
understand multiscale phenomena such as turbulence.

\newcommand{\Kn}{\mbox{Kn}}

Some gas flows are not fluids and \emph{must} be treated by a
kinetic-theory approach.  The nature of the flow is described in
kinetic theory by the ratio of the mean free path to the
characteristic scale, called \emph{Knudsen} number or $Kn$:
\begin{equation}
\mbox{Knudsen number} = \Kn \equiv \frac{\mbox{Mean free
    path}}{\mbox{Characteristic size}} = \frac{\lambda}{L}
\label{eq:kn}
\end{equation}
In astrophysics, typical characteristic scales include density,
temperature, and gravitational-field scale lengths.  For $\Kn>0.05$,
the solutions of the continuum fluid equations deviate from the exact
solutions \citep{Boyd.etal:1995}.  It should not be surprising that
many of the most important astrophysical regimes are near this
boundary, as shown in Figure \ref{fig:regimes}.  The knowledge of the
physical state within the transition regime is necessary for
predicting the physical processes that may dominate the observed
emission or cooling even if \Kn\ is small elsewhere.  In other
regimes, the continuum limit is not appropriate even without a
transition region or discontinuity.

\section{Method}
\label{sec:method}

The Direct Simulation Monte Carlo method \citep[DSMC, see][]{Bird:94}
has been widely used in engineering applications for gas flows beyond
the continuum limit.  DSMC is a numerical method, originally conceived
for modeling rarefied gas particles with mean free paths the same
order or greater than the characteristic physical length scale
(i.e. $\Kn > 1$).  In such rarefied flows, the Navier-Stokes equations
can be inaccurate. But recently, the DSMC method has been extended to
model near continuum flows ($\Kn\ll0.05$), making it appropriate for
astrophysical problems with large dynamic ranges and multiple phases.
In addition, DSMC is fully shock capturing in the sense that the
discontinuity implied by the shock is naturally determined by DSMC.
It reproduces standard shock tests (see \S\ref{sec:shock}) and
instabilities such as Kelvin-Helmholtz (see \S\ref{sec:KH}).  Because
the method works for arbitrary values of \Kn, DSMC may be used in
astrophysical problems that have short and long mean free paths that
might occur in the interaction between cold gas and hot rarefied gas
in cluster environments (e.g. see Paper 2 for an application to
ram-pressure stripping).

Furthermore, DSMC is always stable (no Courant-Friedrichs-Lewy
condition). As will be described in \S\ref{sec:tuning}, there
\emph{are} conditions on the step size and collision parameters for
optimal performance, but poor parameter choices lead to some
inaccuracy but not failure through instability.  The main problem with
DSMC approach is computational speed: for a point of comparison, DSMC
is more efficient per particle than a smoothed-particle hydrodynamics
(SPH) code, but it requires at least an order of magnitude more
particles to achieve a similar resolution.  The performance in our
implementation is bottlenecked by the tree structure supporting the
parallel domain decomposition and thereby marks this area for further
technical development.  Nonetheless, DSMC will never compete in speed
and accuracy with hydrodynamic methods in the $\Kn\rightarrow0$ limit.
Rather, the goal of the DSMC method is an investigation of the effect
of microphysics and gas dynamics in the transition regimes themselves;
these regimes cannot be studied within the hydrodynamic paradigm.

\subsection{Introduction to Direct Simulation Monte Carlo}
\label{sec:DSMC}

DSMC incorporates the internal degrees of freedom of the atoms and
molecules in a gas through a variety of approximations that
redistribute the kinetic energy, momentum and internal energy of two
collision partners.  For a simple monoatomic gas, the only relevant
energies are those of translation, electronic excitation and radiative
emission by the particles.  This system is described by a phase-space
distribution function, $f_k(\mathbf{x}, \mathbf{v}; \epsilon_j, t)$,
that represents the expected number density of molecules of species
$k$ in a small volume $d\mathbf{x}$ about the point at $\mathbf{x}$
which have a velocity in the range $\mathbf{v}$ to $\mathbf{v} +
d\mathbf{v}$, and internal energy $\epsilon_j$ a given instant $t$.

The limiting kinetic equation for DSMC is the nonlinear collisional
Boltzmann equation \citep{Wagner:92}.  DSMC splits the simulated
solution of the collisional Boltzmann equation into two
sequentially-applied parts.  First, the particles are first advanced
on collisionless trajectories using the standard n-body method.
Second, the flow field is divided into cells, and collisions are
between the simulation particles realized consistent with the local
collision rate.  DSMC represents the atoms and molecules of a gas by
much smaller number of simulation particles.  The number of collisions
reproduces the rate in the physical system by increasing the cross
section for the simulation particles by the ratio of the simulation
mass to the true mean atomic and molecular mass.  This generic feature
is exploited in a variety of ways to improve DSMC performance.  For
example, if the collision rate is sufficiently large, the velocity
distribution in a cell will approach the equilibrium Maxwell-Boltzmann
distribution and this may be used to limit the number of computed
collisions explicitly.  If the collision rate is low, however, there
will be little redistribution of energy and momentum and the
distribution will remain close to the collisionless solution which may
be far from thermodynamic equilibrium.  The details of the algorithm
will be outlined in \S\ref{sec:algorithm} below.

DSMC is presently the most widely used numerical algorithm in kinetic
theory \citep{Garcia.Wagner:2000}. In DSMC, particle pairs are
randomly chosen to collide according to the probability distribution
implied by the interparticle potential.  This probability is
proportional to the particles' relative speed and effective geometric
cross section. The post-collision velocities are determined by
randomly selecting the collision angles and redistributing the energy
into atomic and molecular internal degrees of freedom.  Therefore,
unlike molecular dynamics, DSMC particles are chosen to collide even
if their actual trajectories do not overlap.  This is not an
inconsistency but required by the probabilistic nature of the
solution.

The application of the \emph{classic} DSMC algorithm is restricted to
dilute gases.  Recently, the Consistent Boltzmann Algorithm (CBA) was
introduced as a simple variant of DSMC for dense gases
\citep{Alexander.etal:95}. The CBA collision algorithm follows the
DSMC algorithm with two additions. First, the unit vector parallel to
the line connecting the centers at impact is computed from the pre-
and post-collision velocities of the colliding pair. Each particle is
displaced along this direction corresponding to the mean separation
they would have experienced had they collided as hard spheres.
Secondly, the collision rate must be increased over the dilute rate to
account for the volume displaced by the hard spheres representing the
atoms and molecules themselves. These algorithmic changes are unlikely
to play a major role in astrophysical flows, but are trivial to
include.  With these two simple additions, CBA yields the hard sphere
equation of state at all densities.  \citet{Montanero.Santos:96}
showed that the high-density transport properties are not exact but
remain comparable to with molecular dynamics simulations and the
Enskog approximation.  Although CBA can be generalized to any equation
of state here we will only consider the hard sphere gas whose particle
diameter is a constant fraction of the Bohr diameter.

\subsubsection{Details of the DSMC Algorithm}
\label{sec:algorithm}

The two sequential steps that simulate equation (\ref{eq:be}) are a
ballistic or collisionless computation based on the left-hand side and
a collisional computation based on the right hand side.  Practically,
one advances the particles according the collisionless Boltzmann
equation,
\begin{equation}
  \frac{\partial f}{\partial t} + \mathbf{v}\cdot {\mathbf{\nabla}_{\bf
      x} f} + \mathbf{F}\cdot{\mathbf{\nabla}_{\bf v} f} = 0
  \label{eq:advect}
\end{equation}
using standard n-body techniques.  One then solves for the collisions
using
\begin{equation}
  \frac{\partial f}{\partial t} = Q(f, f).
  \label{eq:collide}
\end{equation}
The collision operator is
\begin{eqnarray}
Q(f, f) &=& \int_{\mathbb{R}^3} \int_S
\sigma(|\mathbf{v}-\mathbf{v}_\ast|, \Omega)
|\mathbf{v}-\mathbf{v}_\ast|
\times \nonumber \\
&& 
\left[f(\mathbf{v}^\prime)f(\mathbf{v}^\prime_\ast) -
f(\mathbf{v})f(\mathbf{v}_\ast)\right] \,d\Omega\,d\mathbf{v}_\ast
\label{eq:collop}
\end{eqnarray}
where $\sigma(\cdot,\cdot)$ denotes the collision cross section, the
primed $\mathbf{v}^\prime$ and $\mathbf{v}^\prime_\ast$ describe all
the possible post-collisional velocities of two particles colliding
with respective pre-collisional velocities $\mathbf{v}$ and
$\mathbf{v}_\ast$, and $\Omega$ denotes the interaction angles.  In
DSMC, the collision operator is solved by a Monte Carlo realization of
equation (\ref{eq:collop}) as follows:
\begin{enumerate}
\item Move all of the particles collisionlessly according to the mean
  field (eq. \ref{eq:advect}) using the n-body solver.
\item Partition particles into cells whose linear scale is of order
  the mean free path $\lambda$.
\item Compute collision frequency in a cell for all particle species
  and interactions of interest.  This discretizes the spatial
  dependence of the phase-space density $f$ to provide
  $f(\mathbf{v})$.
\item Select random collision partners within cell; this samples
  $f(\mathbf{v})$ and $f(\mathbf{v}_\ast)$ in equation
  (\ref{eq:collop}).  We assume that probability that a pair collides
  only depends on their relative velocity. That is, all particles in
  the cell are valid collision partners.  The post-collision
  velocities (6 quantities) are constrained by the conservation of
  momentum (3 constraints) and energy (1 constraint).  The
  post-collision direction in the center of mass frame is specified by
  two randomly chosen variates.
\item Repeat these steps beginning with Step (i).
\end{enumerate}
In this way, the net change in the phase-space density implied by
equations (\ref{eq:be}) and (\ref{eq:collop}) are computed by evolving
velocity distributions consistent with the interaction potential
implicit in the cross section $\sigma$.  This, in turn, leads to mass,
momentum, and energy flux through the collision cells.  See
\citet{Bird:94} and \citet{Cercignani:00} for additional practical
details and theoretical underpinning of this approach.

\subsubsection{Algorithm efficiency and tuning}
\label{sec:tuning}

\begin{figure}
\centering
\includegraphics[width=0.5\textwidth]{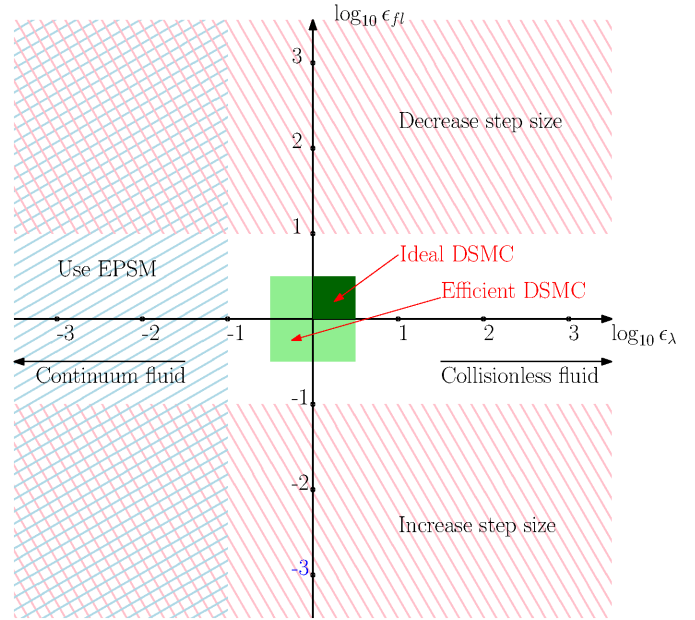}
\caption{Schematic description of DSMC parameters on the
  $\epsilon_{\lambda}$--$\epsilon_{fl}$ plane.  The light green box
  shows the desirable range for the two parameters for efficiency.
  The dark green box shows the ideal range.  If $\epsilon_{fl}$ is too
  large, particles may cover multiple cells for
  $\epsilon_{\lambda}\gta1$ (red shaded region); the step size should
  be reduced.  If it is too small, the particles may interact with
  particles that can not be reached by free flight in one time step.
  For $\epsilon_{\lambda}\lta0.1$ (blue shaded region), the limiting
  EPSM algorithm is used.}
\label{fig:tuning}
\end{figure}

DSMC is most efficient and accurate when the interaction cell size $l$
is of order the mean free path $\lambda$ \citep{Sun.etal:2011} and a
free particle crosses the cell in a time step (${\bar v}\delta
t\approx l$ where ${\bar v} $ is the mean one-dimensional particle
velocity and $\Delta t$ is the time step); let us call ${\bar v}\Delta
t$ he \emph{flight length}.  This motivates defining two scale-free
quantities:
\begin{eqnarray}
  \epsilon_{\lambda} &\equiv& \lambda/l, \label{eq:epsL} \\
  \epsilon_{fl}     &\equiv& {\bar v}\Delta t/l. \label{eq:epsV}
\end{eqnarray}
The mean-free path ratio, $\epsilon_{\lambda}$, is the mean-free path 
$\lambda $ in units of the cell size $l$.  The flight-length ratio,
$\epsilon_{fl}$, is the length of flight in one time step in units of cell
size.  The cell size $l$ is chosen to satisfy two additional
constraints: 1) $l$ should be smaller than any flow scale of interest;
and 2) the cell size should enclose approximately 10 particles.
Clearly, if the cell size is so small that the probability of
occupation is small, no interactions are possible.  Conversely, if the
occupation number is large, the expense may increase without improving
the accuracy of the result.  This parametrization leads to the
following limiting scenarios and tuning prescriptions:
\begin{itemize}
\item $\epsilon_\lambda\sim{\cal O}(1), \epsilon_{fl}\gg1$.  \newline
  Some particles will pass through multiple cells in one time step,
  possibly leading to artificial viscosity (see \S\ref{sec:mfp_eff}).
  The remedy is to decrease step size, $\Delta t$.
\item $\epsilon_\lambda\sim{\cal O}(1), \epsilon_{fl}\ll1$.  \newline
  Particles can not reach their interaction partners in one time step.
  This leads to excess correlation.  The remedy is to increase step
  size, $\Delta t$.  In our DSMC implementation, the appropriate time
  step is selected automatically selected for each gas particle based
  on these criteria as long as the times step criteria demanded by the
  Poisson solver (eqs.  \ref{eq:eps1}--\ref{eq:eps3}) are not
  violated.
\item $\epsilon_\lambda\gg1, \epsilon_{fl}\sim{\cal O}(1)$.  \newline
  Mean free path is very short compared to any scale of interest.  The
  system is approaching the continuum limit with many collisions per
  particle in a time step.  One remedy is to use a collision limiting
  scheme.  Here, we use the Equilibrium Particle Simulation Method
  (EPSM) described in \S\ref{sec:epsm}.  When using EPSM, the relevant
  velocity in equation (\ref{eq:epsV}) is the mean flow velocity
  $\langle v\rangle$ not the mean ballistic velocity ${\bar v}$.  Of
  course, if this condition obtains everywhere in the computational
  volume, all cells will use EPSM and the resulting calculation will
  be in the CFD regime.
\item $\epsilon_\lambda\ll1, \epsilon_{fl}\sim{\cal O}(1)$.  \newline
  Mean free path is very large compared to any scale of interest.  The
  partial remedy is to increase $l$ and possibly $\Delta t$ as long as
  $l$ remains smaller than any scale of interest.  Otherwise, the
  system is approaching the collisionless limit and all is well.
\end{itemize}
This regimes are summarized schematically in Figure \ref{fig:tuning}.

\subsubsection{Continuum--DSMC Hybrids}
\label{sec:epsm}

DSMC is computationally expensive, although much cheaper than
molecular dynamics.  The expense can be mitigated by only using DSMC
where it is needed!  A number of recent contributions to the
literature describe hybrid Navier-Stokes--DSMC solvers using
finite-element methods and AMR techniques
\citet[e.g.][]{Garcia.etal:1999,Wissink.etal:2001a,Wijesinghe.etal:2004}.
Because of the unusual geometries and large dynamic range in
astrophysical flows, the implementation of the kinetic theory is
greatly simplified by using a particle method throughout these
simulations rather than a hybrid DSMC--Navier-Stokes code.  A pure
particle code easily accommodates the arbitrary geometry of
interfaces, simplifies the transition between regimes without the
numerical artifacts owing to the statistical noise incurred by moving
between the particle simulation and continuum representation.
However, in the high density regions that are collision dominated and
make DSMC infeasible, the system will approach thermodynamic
equilibrium.  Based on this, \citet{Bird:2007} suggested that the
number of collisions per particle be limited to only the number
necessary to achieve equilibrium.

A further simplification eliminates individual collision computations
altogether when the density and collision rates are high.  If one can
predict that the number of collisions will be sufficient to achieve
equilibrium without simulating the collisions, the limiting
thermodynamic state for a collision-dominated cell may be computed a
priori, and one may achieve a significant computational advantage.
This high-collision limit of DSMC was proposed by \citet{Pullin:80}
who called it the Equilibrium Particle Simulation Method (EPSM).  This
approach has been more recently explored by \citet{Macrossan:2001}.
In essence, EPSM selects new velocity components from the equilibrium
thermodynamic distribution within each cell while simultaneously
conserving the total energy and momentum.  An estimate of the
temperature, ${\hat T}$ may be derived from the total energy of the
finite sample of $n_c$ particles in a cell using the standard relation
for the mean energy of single degree of freedom $\epsilon_{dof} =
\frac12 k_b {\hat T}$.  Although Pullin's application was a single
species in one dimension, the approach is straightforwardly
generalized to any number of species in three dimensions.  We will
combine the DSMC and EPSM collision computation approximations using a
heuristic such as the mean number of collisions between particles to
switch between the two.  More sophisticated collision limiting
approaches with fewer limitations are under investigation
\citep[see][]{Zhang.etal:2008} and will be explored in the future.

Owing to the random realization of evolved states in both DSMC and
EPSM, the EPSM solution will differ from the DSMC solution and the
exact molecular dynamics solution, even after many collisions.  These
differences lead to statistical scatter about the exact solution.
Presumably, the scatter will be larger for DSMC than for EPSM, since
initially identical initial states will contain slightly different
total amounts of energy and momentum. It might also be possible to
improve the use of both algorithms together by conserving or matching
incoming and outgoing fluxes at cell boundaries between the two
regimes.

EPSM is expected to be more efficient than DSMC close to thermodynamic
equilibrium.  However in DSMC, the time step is chosen to be of order
the collision time.  Obviously, this condition must be relaxed in
EPSM, otherwise the mean number of collisions per particle would be
${\cal O}(1)$ and the distribution could be far from equilibrium.  In
both schemes, large cell sizes $\delta x$ and step sizes $\Delta t$
larger than the local collision time may be used in regions where the
flow gradients are small.  See \S\ref{sec:mfp_eff} for additional
complications.

\subsubsection{Effective mean free path and dissipation in DSMC}
\label{sec:mfp_eff}

To summarize, EPSM is an infinite collision rate limit of DSMC with a
finite sample of simulator particles.  If the time step exceeds the
time of flight across a typical cell, or equivalently $\epsilon_{fl}>1$, 
then particles with momentum and energy representative of the equilibrium
distribution of one particular cell can be \emph{non-physically}
transported to a distant cell with possibly different equilibrium
conditions.  Close to equilibrium, this distance is approximately
$(k_BT/m)^{1/2}\Delta $, where $m$ is the particle mass.  On the other
hand, the distance covered by particles in $\Delta t$ should be at
least as large as the mean free path to prevent inducing non-physical
correlations.  If the criteria described in \S\ref{sec:tuning} are
close to ideal, the probability that particles will cross the cell in
any one time step is small in the high-collision-rate limit.  However,
the ideal tuning parameters are not always computationally feasible.

To understand the implications of these limits, let us consider
transport of particles between two adjacent cells.  Consider a
Euclidean set of coordinate axes for a collision cell, $({\hat x},
{\hat y}, {\hat z})$ with ${\hat x}$ chosen perpendicular to the cell
face common to the adjacent cells.  Let this face be located at $x=0$.
Let the mass density of the two cells to the left and right of the
boundary $x=0$ be denoted by $\rho^{<0}$ and $\rho^{>0}$,
respectively.  For notational simplicity, let factor the phase-space
distribution function into a density and velocity distribution
function: $f(\mathbf{x}, \mathbf{v}) = \rho(\mathbf{x})
g(\mathbf{v})$; this is consistent with our discretization of the
phase space into collision cells.  The velocity distribution function
describing the state in the adjacent cells will be different in
general: $g^{<0} \not= g^{>0}$.  That is, particles that cross the
cell boundary with $v_x>0$ ($v_x<0$) have the parent distribution
function $g^{<0}$ ($g^{>0}$).  This discontinuity gives rise to an
effective viscosity.

\newcommand{\Flux}{{\cal F}}

Demonstrating this explicitly is an easy exercise using the standard
techniques from kinetic theory \citep[e.g.][]{Cercignani:90}. Let $q$
denote some velocity moment of $g$, such as the momentum.  The net
flux of the $q$ across the interface, $\Flux_q$, can be split into two
contributions by the direction of the particles crossing the cell
face, $\Flux_q = \Flux_q^{<0} + \Flux_q^{>0}$, as follows:
\begin{eqnarray}
  \Flux_q^{<0} &=& \frac{\rho^{<0}}{m} \int^{\infty}_{0} dv_x
  \int^{\infty}_{-\infty} dv_y\int^{\infty}_{-\infty} dv_z
  g^{<0} v_x q({\bf v}), \\
  \Flux_q^{>0} &=& \frac{\rho^{>0}}{m} \int^{0}_{-\infty} dv_x
  \int^{\infty}_{-\infty} dv_y\int^{\infty}_{-\infty} dv_z
  g^{>0} v_x q({\bf v}).
\end{eqnarray} 
In general, the gradient in velocity parallel to cell wall will not
vanish across the interface: $\partial v_y/\partial x \not=0$.
Setting $q=mv_y$ and taking $g^{<0}$ and $g^{>0}$ to be
Maxwell-Boltzmann distributions with the temperatures $T^{<0}$ and
$T^{>0}$ in the adjacent cells, the momentum fluxes parallel to the
cell wall are
\begin{eqnarray}
  \Flux^{<0}_{mv_y} &=& 
  \frac12\left[1 + 
    \erf\left(\langle v_x^{<0}\rangle/\sqrt{2k_bT^{<0}/m}\right) \right] 
  \rho^{<0} \langle v_x^{<0}\rangle\langle v_y^{<0}\rangle \nonumber \\ && + 
  \sqrt{\frac{k_bT^{<0}}{2\pi m}} e^{-\langle
    v_x^{<0}\rangle^2/2k_bT^{<0}/m}  \rho^{<0} \langle v_y^{<0}\rangle \\
  \Flux^{>0}_{mv_y} &=& 
  \frac12\left[1 - \erf\left(\langle
      v_x^{>0}\rangle/\sqrt{2k_bT^{>0}/m}\right)\right]
  \rho^{>0} \langle v_x^{>0}\rangle\langle v_y^{>0}\rangle  
  \nonumber \\ && - 
  \sqrt{\frac{k_bT^{>0}}{2\pi m}} e^{-\langle
    v_x^{>0}\rangle^2/2k_bT^{>0}/m} \rho^{>0} \langle v_y^{>0}\rangle  
  \\
\end{eqnarray}
where
\begin{equation}
  \langle v_j^{<0} \rangle = 
  \int^{\infty}_{-\infty} dv_x 
  \int^{\infty}_{-\infty} dv_y
  \int^{\infty}_{-\infty} dv_z
  g^{<0} v_j^{<0}
\end{equation}
and analogously for $\langle v_j^{>0} \rangle$.  Assume that there are
no additional gradients for simplicity; this implies that $T\equiv
T^{<0} = T^{>0}$ and $\rho\equiv\rho^{<0}=\rho^{>0}$.  In addition,
assume that the mean velocity is small compared to the thermal
velocity, $\langle v_j^{>0} \rangle \ll \sqrt{k_bT/m}$.  Then, the net
momentum flux across the cell boundary becomes
\begin{equation}
  \Flux_{mv_y} = \rho \sqrt{\frac{k_bT}{2\pi m}} \left(\langle
    v^{>0}_y\rangle - \langle v^{>0}_y\rangle\right) + {\cal
    O}(\langle v_j\rangle^2).
\end{equation}

Now, define the distance that the mean particle transports its
momentum as $s$.  The value of $s$ will depend both the details of
DSMC code and the physics of the interactions and gives us an
approximation to the gradient:
\begin{equation*}
  \frac{\partial v_y}{\partial x} \approx \frac{\langle v_y^{>0}\rangle -
    \langle v_y^{<0}\rangle}{s}.
\end{equation*}
Combining these relations, we may express the net momentum flux as
\begin{equation}
  \Flux_{mv_y} \approx \rho \sqrt{\frac{k_bT}{2\pi m}}
  \frac{\partial v_y}{\partial x} s.
  \label{eq:stress}
\end{equation}
This is a shear stress induced at the cell boundary by the
discretization.

Physically, viscosity arises from the shear stress at an interface
that opposes an applied force.  The classic example is the laminar
flow of a viscous fluid in the space between two relatively moving
parallel plates, known as Couette flow.  The force applied to the
plates causes the fluid between the plates to shear with a velocity
gradient in the direction of relative motion.  In other words, the
shear stress between layers is proportional to the velocity gradient
in the direction perpendicular to the layers:
\begin{equation}
  \tau=\mu \frac{\partial v}{\partial x}
  \label{eq:couette}
\end{equation}
where $\mu$ is the proportionality factor called the \emph{viscosity}.

Comparing equation (\ref{eq:stress}) that describes our discretization
shear stress to the usual relation between mean free path and
viscosity for planar Couette flow (eq. \ref{eq:couette}), we may
estimate the effective viscosity for the simulation in the near
equilibrium limit to be
\begin{equation}
  \mu_{sim} = \rho\sqrt{\frac{k_BT}{2\pi m}}\,s
  \label{eq:effvisc}
\end{equation}
The mean free path for a hard sphere with viscosity $\mu$ is
\begin{equation}
  \lambda_v = \frac{\mu}{\rho}\left(\frac{\pi m}{2 k_B T}\right)^{1/2}
  \label{eq:hsvisc}
\end{equation}
\citep{Cercignani:90}.  This implies that the effective mean free path
due to the transport across the cell boundary in the simulation is
then $\lambda_{sim} = s/2$.

The EPSM algorithm mixes the transported momentum into the entire cell
of size $l$.  Interestingly, this suggests that $s=l$ and the
effective mean free path in DSMC simulation becomes
\begin{equation}
  \lambda_{DSMC} = \max\{l/2, \lambda\}
\end{equation}
where $\lambda$ is the local mean free path.  This is equivalent to
the DSMC requirement that the cell size should be smaller than the
collisional mean free path for physical accuracy.  However, when scale
of all macroscopic observables is large compared to the mean free
path, the less expensive EPSM solution should closely approximate the
DSMC solution.  In other words, DSMC and EPSM should give similar
results when the number of collisions per particle per time step in
DSMC is large, which justifies the use of cell sizes much larger than
the mean free path and artificially limiting the number of collisions
per particle in this limit.

Similarly, the same arguments leading to equation (\ref{eq:stress})
suggest that if DSMC grid cells are too large ($l\gg\lambda$),
non-local particles will be selected for collisions resulting in the
numerical diffusion of gradients on the cell-size scale. Conversely,
if DSMC cells are too small ($l\ll\lambda$), but still contain the
same number of particle per cell, the number of simulated particles
becomes far more than necessary, resulting in an accurate but highly
inefficient simulation.

In some flow problems of astrophysical interest, it is not possible to
use a sufficient number of particles to populate cells at the
mean-free-path scale.  In these cases, we resort to using an
under-resolved collision-limited DSMC solution with large cells.  As
we have seen, the overall effect of such an approximation is to
misrepresent the transport coefficients giving rise to an effective
viscosity.  This should not lead to significant misestimations as long
as the gradients of the flow are small with length scales greater than
the cell size.

\subsection{Implementation}
\label{sec:implement}

\subsubsection{N-body code}
\label{sec:nbody}

We have implemented the DSMC kinetic algorithm in our n-body expansion
code, EXP.  Owing to its long effective relaxation times, the
expansion or ``self-consistent field'' algorithm (SCF, e.g.  Hernquist
\& Ostriker 1992\nocite{Hernquist.Ostriker:92}) is ideal for
representing the large-scale global response of a galaxy to a
long-term disturbance.  Weinberg (1999\nocite{Weinberg:99}) extended
this algorithm from fixed analytic bases to adaptive bases using
techniques from statistical density estimation to derive an optimal
smoothing algorithm for SCF that in essence selects the minimum
statistically significant length scale.  This is combined with the
empirically determined orthogonal functions that best represent the
particle distribution and separate any correlated global patterns.
The rapid convergence of the expansion series using this matched basis
minimizes fluctuations in the gravitational force by reducing the
number of degrees of freedom in the representation and by increasing
the signal-to-noise ratio for those that do contribute.  In addition,
EXP contains a particle-in-cell Poisson solver as well as a
parallelized direct-summation solver.  The former may be ideal for
investigating the importance of small self-gravitating spatial
inhomogeneities that are a small part larger astrophysical flow.

The EXP code is modular with an object-oriented architecture and
easily incorporates DSMC.  The standard SCF algorithm is
``embarrassingly'' parallel and the implementation here uses the
Message Passing Interface communications package (MPI,
e.g. \citealt{Gropp.etal:1999a,Gabriel.etal:2004}), making the code
easily portable to a variety of parallel systems.  Practically
speaking, parallel SCF with the new algorithm makes tractable disk and
halo simulations on modest-sized PC-based clusters with up to 100
million particles.  

EXP uses a multiple time step algorithm as follows.  We begin by
partitioning phase space $m+1$ ways such that each partition contains
$n_j$ particles that require a time step
\begin{equation}
  \delta t_j = 2^{-j}\delta T,  \quad\mbox{with}\quad j=0,\ldots,m
  \label{eq:ts}
\end{equation}
where $\delta T$ is the largest time step.  The time step with $j=m$
corresponds to the smallest single time-step in the simulation.  Since
the total cost of a time step is proportional to the number of force
evaluations, this algorithm improves the run time by the following
factor
\begin{eqnarray}
{\cal S} &=& \sum_{j=0}^{m}n_j 2^j/\sum_{j=0}^{m} n_j 2^{m}
\nonumber \\ &=&
\sum_{j=0}^{m}n_j 2^{j-m+1}/\sum_{j=0}^{m} n_j = 
\frac{1}{N}\sum_{j=0}^{m}n_j 2^{j-m+1}
\end{eqnarray}
where $N=\sum_{j=0}^{m}n_j$.  If all particles were the deepest level,
$j=m$, we have ${\cal S}=1$.  On the other hand, if most of the
particles are in the level $j=0$, we have ${\cal S}=2^{m}$.  For an
$c=15$ NFW dark-matter profile \citep{Navarro.Frenk.ea:97} with
$N=10^7$ particles as an example, we find that $m=8$ and ${\cal
  S}\approx 30$, an enormous speed up!  Forces in the SCF algorithm
depend on the expansion coefficients and the leap frog algorithm
requires interpolation of these coefficients to maintain second-order
error accuracy per step.  The contribution to each expansion
coefficient for particles in time-step partition $j$ are separately
accumulated and linearly interpolated for levels $k>j$ as needed.
Higher-order interpolation would have higher-order truncation error
than the ordinary differential equation solver and would be wasteful.

EXP uses the minimum of three separate time step criteria for choosing
the appropriate time-step partition $j$ for each particle $i$:
\begin{enumerate}
\item A local \emph{force} time scale:
  \begin{equation}
  \Delta t_1 = \epsilon_1 \frac{|\mathbf{v}_i|}{|\mathbf{\nabla\Phi}_i|}.
  \label{eq:eps1}
  \end{equation}
  This sets the time step to be a fraction $\epsilon_1$ of the
  estimated change of local velocity by the gravitational force; that
  is, this time step is based on the rate of velocity change implied
  by the gravitational field.
\item A local \emph{work} time scale: 
  \begin{equation}
  \Delta t_2 = \epsilon_2 \frac{|\Phi_i|}{|\mathbf{v}_i\cdot\nabla\Phi_i|}.
  \label{eq:eps2}
  \end{equation}
  This sets the time step to be a fraction $\epsilon_2$ of the time
  required to change the particle's potential energy.
\item And a local \emph{escape} time scale:
  \begin{equation}
  \Delta t_3 = \epsilon_3 \sqrt{\frac{|\Phi_i|}{|\nabla\Phi_i|^2}}.
  \label{eq:eps3}
  \end{equation}
  This time step will be small for a fast moving particle near a local
  feature in the gravitational field or for a particle that samples a
  large range of gravitational field strengths over its orbit.
\end{enumerate}
We calibrate the leading coefficients $\epsilon_k$ for accuracy using
a test simulation for each new set of initial conditions.  Typically
$\epsilon_k\approx0.01$ yields better than 0.3\% accuracy over a
dynamical time.

\subsubsection{DSMC parallelization}
\label{sec:parallel}

Our implementation uses an octree for partitioning space to make DSMC
collisions cells.  Each node in the octree is a rectangular prism most
often chosen to be a cube.  Each node is dividing into eight child
nodes of equal volume formed by a single bisection in each dimension.
Octrees are the three-dimensional analog of quadtrees.  Unlike
kd-trees, the octree partitioning is on volume alone, so that the
aspect ratio of all nodes are self similar.  This is desirable for
collisions cells that assume that any particle in a cell whose linear
dimension is approximately a mean free path is a good collision
partner for another particle from that same cell.

Most contemporary supercomputers are networked clusters with multiple
multi-core processors per node.  Each of the CPU cores share the
memory and communication resources of the parent node, and each node
is interconnected to the others through a fast low-latency network.
To simplify the parallel domain decomposition and exploit the typical
cluster topology, we use a two-level nested tree as follows.  The
first level is a coarse octree constructed with approximately an order
of magnitude more cells than nodes.  The computational work for all
particles belonging to leaves of the coarse tree is accumulated and
used to balance load among the nodes when the domain is decomposed.
The second level constructs a tree in each leaf of the coarse
first-level tree. Spatially contiguous leaves are preferentially
assigned to the same node.  The goal of the second-level tree are
collections of interaction cells that contain approximately $n_c=10$
simulation particles.  Aggregated cells or super cells used for
statistical diagnostics contain $N_c>n_c$ particles, typically
$N_c\approx 10n_c$.  To minimize the intranode communication overhead,
we run one multithreaded process per node.  Multithreaded tasks
include determining expansion coefficients, force evaluation, and DSMC
collision computations.

The full domain decomposition is computationally expensive and would
horribly dominate the run time if performed at the smallest time step:
$j=m$ in equation (\ref{eq:ts}).  Rather, the tree is extended as
necessary by updating the particle-node association and constructing
new tree nodes as necessary for time intervals short compared to the
evolution time scale but long compared $\delta t_m$.  The interim
procedure updates the octree consistently but without exchanging
particles between processors. At some preselected intermediate time
step assigned by choosing $m_{adj}$ with $0\le m_{adj}\le m$, the
domain decomposition is reconstructed anew, exchanging particles
between processors as necessary. In other words, at time steps with
$j\le m_{adj}$, both the coarse- and fine-grained octrees are
recomputed from scratch.  At time steps with $j > m_{adj}$, the
fine-grained octrees are updated only.  It is therefore possible to
have the \emph{same} cell populated by \emph{different} particles on
separate nodes for a short period of time during the simulation.  This
will introduce some errors since the densities in the duplicated cells
will be underestimated.  However, the value of $m_{adj}$ is chosen so
that $\delta t_{adj}$ remains smaller than any characteristic
dynamical time except for the collision time, and therefore, I do not
anticipate that this procedure will lead to significant quantitative
or qualitative errors.  If there is any doubt, the value of $m_{adj}$
may be increased to $m$ as check.  Along the same lines, DSMC and
DSMC-EPSM explicitly conserve energy and momentum at the collision
scale, including those describing atomic and molecular internal
degrees of freedom.  Therefore, even if small-scale features are
mildly compromised by duplication errors, these are unlikely to
propagate.

\subsubsection{Requirements for accurate DSMC simulations}
\label{sec:accuracy}

Putting together the requirements on collision cells scales and time
steps described in \S\ref{sec:tuning} and \S\ref{sec:mfp_eff}, we may
summarise the interaction between particle number, resolution, cell
size and time step as follows:
\begin{enumerate}
\item The time step must be sufficiently small that the DSMC particles
  do not move through more than one cell of length $l$ in one time
  step, $\Delta T$.  That is the flight-length ratio $\epsilon_{fl}$
  should be of order unity.  The time-step criterion becomes $\Delta T
  \lta l/{\bar v}$.
\item The fundamental resolution scale of the fluid is determined by
  the mean-free path, $\lambda$.  This suggests that collision-cell
  size scale, $l$, should be of order $\lambda$.  Increasing $l$
  beyond $\lambda$ may lead to viscosity as described in
  \S\ref{sec:mfp_eff}.
\item Large separations between collision partners when $\lambda$ is
  large also reduces spatial accuracy. \citet{Bird.etal:09} proposed
  defining virtual \emph{subcells} within collision cells to maintain
  finer spatial control by choosing neighboring collision partners.
  This will be implemented and explored in a future version of our
  code.
\item Each simulation particle represents many atoms and molecules and
  the variance per cell will scale as inversely with the number of
  particles per cell.  However, the computational effort scales as
  $N^2$ so large values of $N$ lead to infeasible simulations.
  Empirically, the overall accuracy scales as $1/N$, and many tests
  suggest that $N\approx 10$--$20$ is a good compromise between these
  competing demands \citep{Fallavollita.etal:1993,Chen.Boyd:1996}.
\end{enumerate}
In summary, the physical properties of the simulation determine the
number of particles required for an accurate DSMC simulation.  Too few
particles degrade the resolution, since collision cells must have $l$
such that $N\gta10$.  Values $l\gg\lambda$ may lead to artificial
viscosity if the characteristic scales of flow gradients are not
resolved.  However, if the mean-free path is smaller than any scale of
interest and the energetics \emph{in} shocks are not critical, the $l$
may be increased artificially without sacrificing reality.  Some
knowledge of the consequences will help motivate whether computational
feasibility justifies this trade off physically.  For example, the
flows on very large scales are likely to be correct owing to accurate
energy and momentum conservation in DSMC independent of $\lambda$ as
long as the small scale features are not critical to determining
features on large scales.  Unlike traditional CFD or N-body simulations, 
a DSMC simulation does not converge with increasing particle number; 
rather, increasing the particle number increases the resolution.  A 
``converged'' solution is obtained by averaging the mean quantities 
from multiple simulations.

\subsubsection{Cooling and heating}
\label{sec:coolheat}

The Monte Carlo realization of the collision integral depends on the
interaction cross section from equation (\ref{eq:collop}) and defines
the probability of internal excitation, ionization, recombination,
radiation or scattering.  Since the spontaneous emission times are
most often very small compared to the collision times for
astrophysical flows, therefore we do not need track the excitation
state of the atoms and molecules between collisions.  Thus,
collisional cooling is a straightforward natural by-product of DSMC.

However, there are some intrinsic difficulties in doing this
accurately that require some additional specialized methods.  For
example, the existence of trace ionized components (e.g. the elements
C, N, O) that have a significant effect on cooling but carry
negligible momentum can not be simulated naively if there will be
fewer than one trace-species particle per interaction cell.  We elect
therefore to change the ratio of the true particle number to
simulation particle number for trace species.  Then, we populate the
simulation with sufficient numbers of the trace species to yield
accurate statistical representation of the trace interactions with
weights to account for their true mass fraction.

A further issue is the treatment of free electrons.  In principle, it
is not difficult to implement free electrons as a separate species.
However as a consequence, the large velocities of the electrons
relative to the ions in equipartition require very small time steps
and lead to infeasibly large computational expense.  We will approach
this problem in two incremental steps.  In the first, we will assume
that the electrons remain in the vicinity of their parent ions, both
preventing charge separation and eliminating the need for time steps
on order of the mean flight time of the simulated electrons.  In the
second, we will include the free electrons as a separate species to
feel the electrostatic field induced by charge separation.
Computational requirements will restrict this application to very
small simulation regions and will be used, primarily, to explore
mesoscale dynamics in various astrophysical regimes.

Alternatively, one may account for the energetic affects of the trace
species by using their reaction rates computed from the thermodynamic
properties approximated from the DSMC fields.  In particular, the
electronic level populations due to collisional and radiative
excitation may be described by rate equations using the electron
density, temperature and background radiation field.  Even if the flow
is far from equilibrium, as long as the processes that establish the
electronic level populations occur much more quickly than the time for
any significant flow pattern to change, an equilibrium distribution
will be a fair approximation in many situations.  This approach is
called the \emph{quasi steady state} (QSS) approximation \citep{Park:1990}.
The electron densities, temperatures and ion densities may then be
used to evaluate the QSS rates.  The QSS rates can be obtained from
individual particle cross sections by assuming that the electron
velocity distributions are Maxwell-Boltzmann.  These same set of
assumptions yields the standard cooling curves used in cosmological
and ISM continuum gas simulations \citep[e.g.][]{Black:81}.  Heating
by cosmic rays, photoelectric heating, etc. may be computed similarly
using the QSS method.

In this and Paper 2, we will employ such an approximate QSS solution
based on LTE.  Then to treat the electrons, we estimate the ionization
fraction based on the local thermal state and assume that the
electrons follow the ions in space.  The implementation here computes
the effective temperature from the super cell of $N_c$ particles (see
\S\ref{sec:parallel}) and estimates collision rates based on a total
effective hard-sphere geometric cross section.

\subsubsection{EPSM implementation}

For very small mean free paths, our algorithm transitions to the
Navier-Stokes equations using the equilibrium particle simulation
through the use of EPSM (\S\ref{sec:epsm}).  If the number of mean
collisions per body in a cell exceeds some preselected value,
$n_{coll}$ The EPSM update is performed in one of two ways.  First,
using the original \citet{Pullin:80} algorithm.  This algorithm
constructively computes random variates while simultaneously enforcing
the constraints of momentum and energy conservation, much the same way
as in the proof for distribution of the consistency of the sample
variance \citep[e.g.][]{Kendall.Stuart:83}.  Alternatively, it is
straightforward to realize normal variates with any convenient mean
and variance, followed by a shift and scale operation to recover the
conserved total momentum and energy.  Both algorithms were implemented
for completeness and yielded equivalent values but the latter is
faster in tests and is the default.  The use of EPSM may be toggled by
a run-time parameter.  In addition, if EPSM is not used, one may elect
to limit the total number of collisions per cell to a maximum value as
suggested by \citet{Bird:2007} while maintaining the correct heating
and cooling rates consistent with the predicted number of collisions.

\subsubsection{Run-time diagnostics}

The simulation particles directly represent the physical properties of
the gas.  They are not \emph{tracers} of an underlying field but
rather all momentum, kinetic energy, internal energy and chemical
fluxes must be computed directly from the particle distribution.
Therefore, the density, temperature and other traditional field
quantities may only be \emph{estimated} as an ensemble average.  Owing
to the typically small number of particles, $n_c$, in an interaction
cell, these estimates are computed using the super cells with $N_c$
particles (see \S\ref{sec:parallel}).  As a consequence, the estimates
used for producing a field representation for the gas' physical state
have lower spatial resolution than the simulation but are useful
diagnostics nonetheless.  In contrast, graphical representations from
hydrodynamic simulations represent the full computed resolution of the
field quantities.  Our current implementation computes ensemble
temperature, density, Knudsen number, and cell size to flight-length
ratio.  These quantities are carried by each gas simulation particle
that are saved in phase-space output files.

The EXP-DSMC module keeps copious diagnostics on the physical
parameters necessary to verify the validity of the DSMC approximation.
For example, the time of flight across the cell, the mean-free path,
and energy radiation rates per cell to ensure that the conditions
required for an accurate DSMC simulation are maintained (see
\S\ref{sec:tuning}).  The necessary time step required for the
particles in each cell is fed back into the time step selection
algorithm to adaptively change the stem steps for particles as
described in \S\ref{sec:nbody}.

\section{Tests}
\label{sec:tests}

\subsection{Shock tube}
\label{sec:shock}

The Riemann shock tube \citep{Sod:78} is a special case of a Riemann
problem in Eulerian hydrodynamics and is defined by an initial state
with two fluids of different density and pressure at rest divided by a
planar interface.  The Rankine-Hugoniot conditions allow one to
compute the flow properties across the shock.  Using this, an exact
solution may be obtained analytically for an adiabatic gas from the
one-dimensional Euler equations written in conservation form.  See
\citet{Bodenheimer.etal:2006} for details. Because an exact solution
is straightforwardly computed, the Riemann shock tube has become a
standard benchmark for computational fluid dynamics problems.  The
shock tube simulation is a strong test for hydrodynamics solvers which
must explicitly and stably compute the shock-front and contact
discontinuities.  Failure may lead to post-shock oscillation in the
solution.

\begin{figure}
  \subfigure[Sod initial conditions, time=0.1 s]{
    \includegraphics[width=0.49\textwidth]{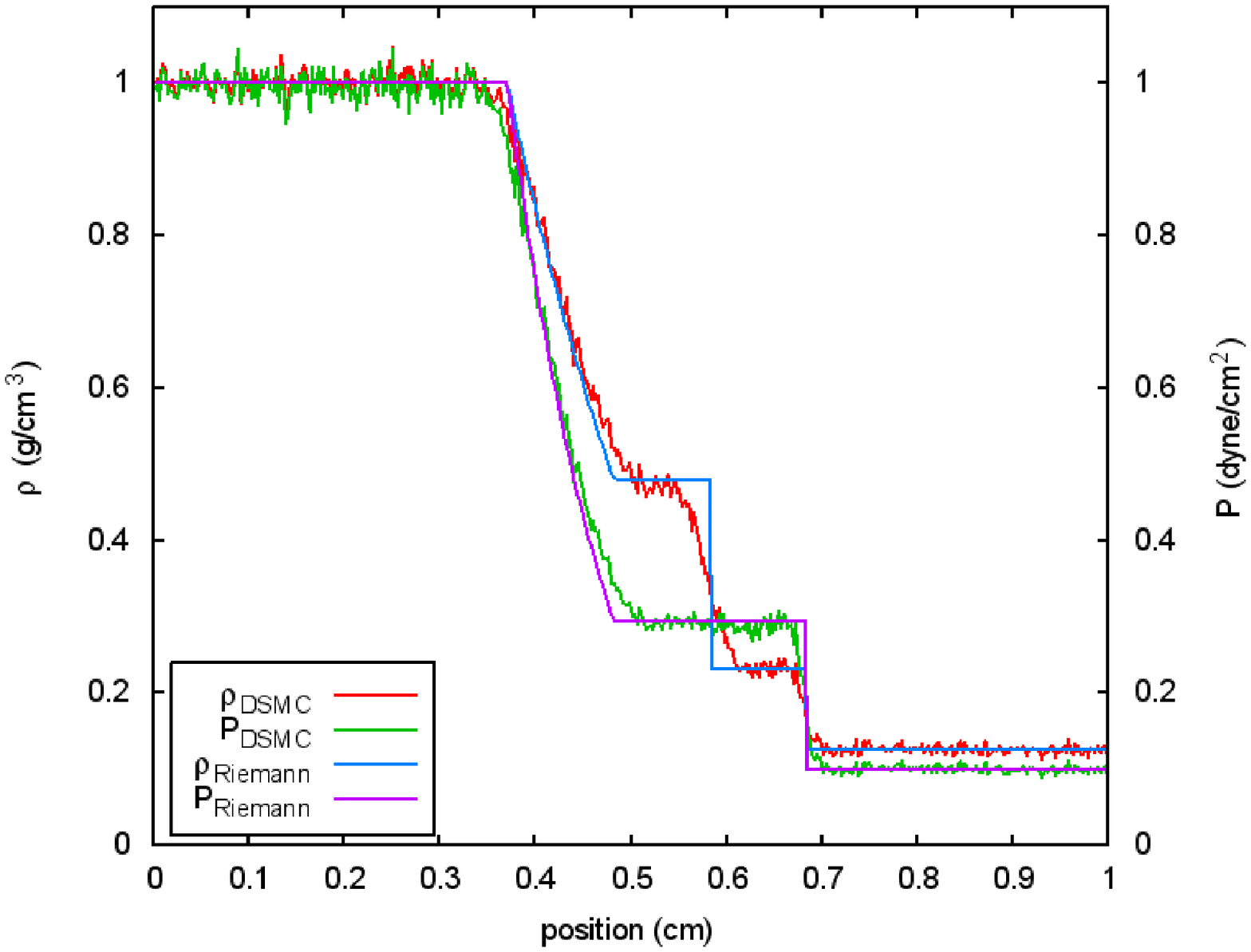}
  }
  \subfigure[Strong initial conditions, time=0.06 s]{
    \includegraphics[width=0.49\textwidth]{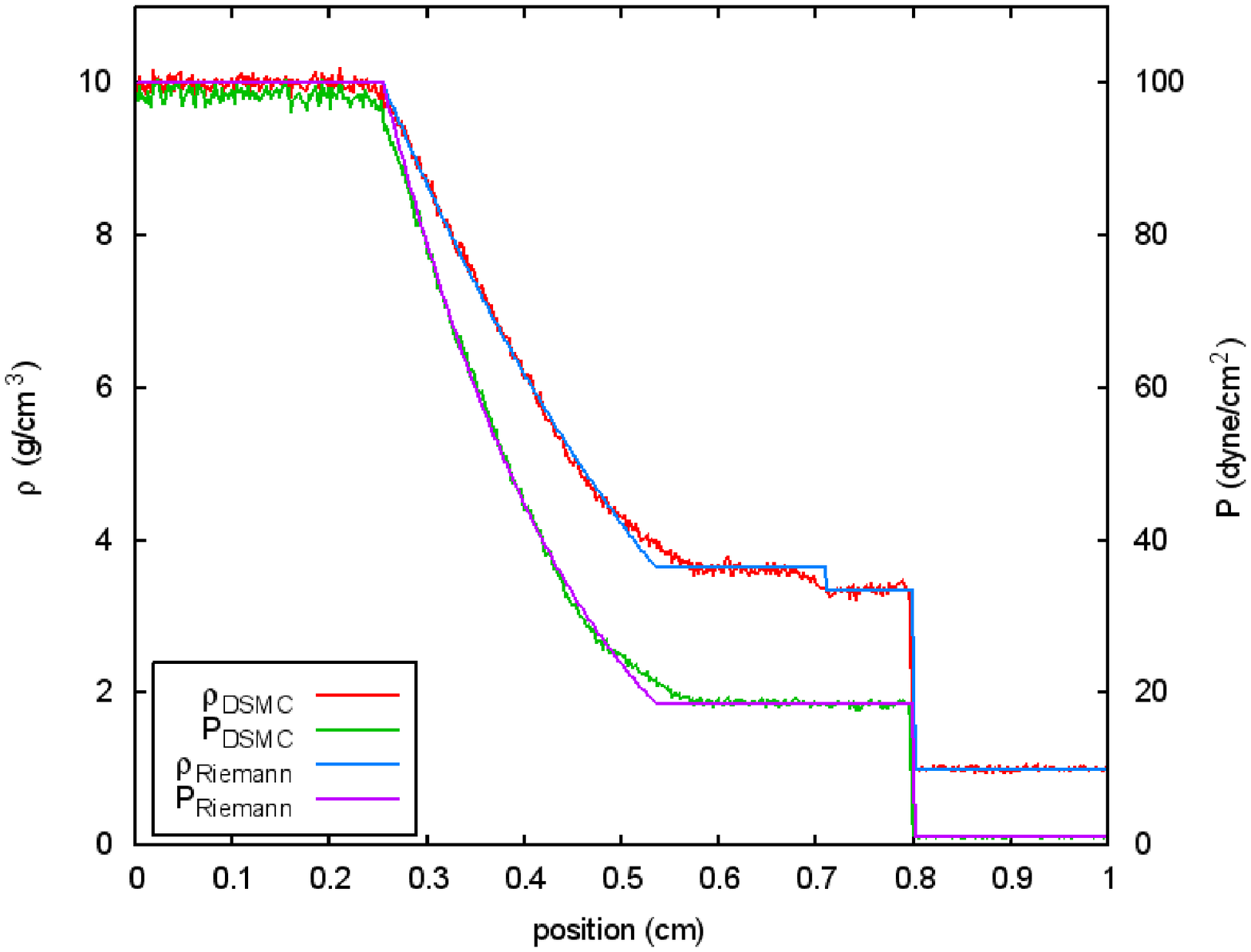}
  }
  \caption{Shock tube test for the classic Sod initial conditions and
    a harder, strong shock (as labeled).  The density and pressure
    axis are shown at the left and right, respectively.  The Regions
    1--5 described in the text are easily identified.  The key
    identifies the numerical DSMC solutions and the exact Riemann
    solver results for both the original and strong-shock initial
    conditions.}
  \label{fig:st1}
\end{figure}

Here, we perform four tests for two different sets of initial
conditions.  The first set of initial conditions is the original Sod
example with $(\rho, P, v)_L=(1,1,0), (\rho, P, v)_R=(1/8,1/10,0)$ and
the second set is a \emph{strong} shock case with $(\rho, P,
v)_L=(10,100,0), (\rho, P, v)_R=(1,1,0)$ suggested by
F. X. Timmes\footnote{\url{http://cococubed.asu.edu/code_pages/ppm_1d.shtml}}.
For each initial condition, we tested a pure DSMC and the hybrid
EPSM/DSMC with indistinguishable results.  Figure \ref{fig:st1}
compares the results of the pure DSMC simulation and the exact Riemann
solution for both cases \citep{Toro:1999}.  The mean-free path is
$\lambda\approx 0.01$ for the Sod initial conditions and
$\lambda\approx 0.005$ for the strong-shock initial conditions,
determined by the particle numbers and computational efficiency as
described in \S\ref{sec:accuracy}.  We expect that the shock and
contact discontinuities will be smeared over several mean-free paths
owing to the internal kinetics of the gas-particle interactions and
some positional variance in the interfaces.  The reproduction of the
density regimes by the DSMC and the DSMC-EPSM hybrid methods without
issues of stability or oscillation is expected owing to the local
nature of the momentum and energy transport in a kinetic simulation.

\subsection{Kelvin Helmholtz instability}
\label{sec:KH}

The traditional Kelvin-Helmholtz (KH) instability is defined for the
flow of two incompressible inviscid fluids with an infinite
plane-parallel interface.  Each fluid has different bulk velocities
$v_1$ and $v_2$ parallel to the interface between the two fluids with
densities $\rho_1$ and $ \rho_2$.  We assume that $v_1>v_2$.  For ease
of discussion, we observe the fluid from the frame of reference moving
at $(v_1+v_2)/2$; this implies that Fluid 1 moves to the right with
velocity $v_1-v_2$ and Fluid 2 moves to the left with velocity
$v_2-v_1$.  For this initial condition, the vorticity is only non-zero
at the interface between the two fluids.  With this initial condition,
an external sinusoidal perturbation causes a growing instability as
follows.  Consider a sinusoidal perturbation to the interface.  The
pressure increases in the concave regions and decreases in convex
regions of the interface (i.e. the Bernoulli principle) which allows
the perturbation to grow.  The peak of the vortex sheet is carried
forward by Fluid 1 and the trough is carried backward by Fluid 2.
This causes the initially sinusoidal corrugation interface to stretch,
tighten and eventually roll up with the same sense as the vorticity at
the original interface.

In Nature, such interfaces abound and the KH instability is thought to
be critical for understanding a wide variety of astrophysical
phenomena.  For example, the non-linear development of the KH
instability leads to turbulent mixing in a “free shear layer”.  These
same interfaces are critical in understanding the development of jets
in radio galaxies and quasars
\citep{Begelman.etal:80,Birkinshaw:91,Perucho.Lobanov:07}.

\begin{figure*}
  \centering
  \includegraphics[width=0.8\textwidth]{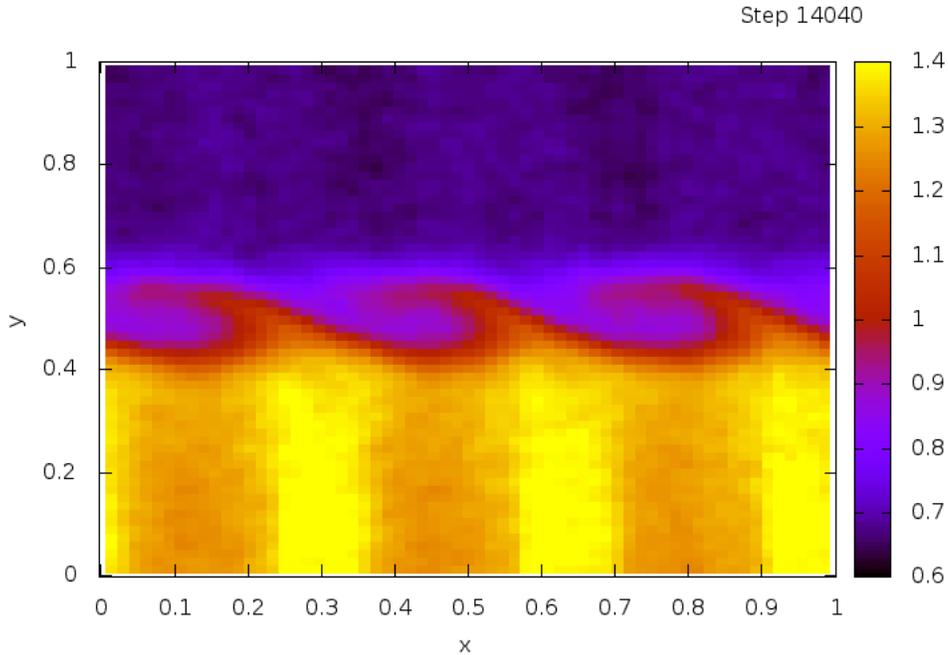}
  \caption{A density slice in the non-linear development of a
    Kelvin-Helmholtz instability seeded with a wave with spatial
    frequency of 1/3.  The sound wave resulting from the seeding is
    clearly visible above and below the interface.}
  \label{fig:kh}
\end{figure*}

Figure \ref{fig:kh} shows the non-linear development of a KH
instability for a fluid box with periodic boundary conditions in the
$x$ and $z$ directions and reflecting boundary conditions in the $y$
direction.  The simulation has $10^6$ particles and unit dimensions
and temperatures $T_1=5000$ K and $T_2=10000$ K.  The density ratio of
the shearing fluids is 2 with pressure equilibrium at the boundary,
$T_1\rho_1 = T_2\rho_2$.  The relative fluid velocity has Mach number
1/2.  The disturbance is seeded everywhere in the box with a
transverse velocity amplitude of 1/4 the shear velocity and spatial
frequency of 1/3 according to the Kelvin-Helmholtz dispersion relation
\citep{Chandrasekhar:1961}.  The instability develops and evolves as
expected, consistent with the spatial frequency and velocity of the
linear mode.  The figure illustrates that the interface is
well-maintained throughout the evolution.  The sound wave that results
from the linear-mode seeded in the initial conditions can be clearly
seen throughout the box.  Of course, the particle numbers place a 
limitation on the maximum spatial frequency that may be resolved owing 
to the DSMC requirements on cell size (see \ref{sec:accuracy}).

\section{Discussion and summary}
\label{sec:summary}

Nearly all gas dynamical simulations on galaxy scales or larger are
numerical solutions of the Navier-Stokes equations (also known as
CFD).  Computational expediency has motivated the use of CFD even when
the mean free path of particles becomes appreciable to the scales of
interest, such as in the early phases of galaxy cluster formation or
in the coronal--neutral gas interface at large galactocentric radii.
In the transition between these regimes and in the rarefied regime
where the mean free paths are of order or larger than scales of
interest, the standard fluid approximation breaks down and one must
solve the Boltzmann equation with collisions using the kinetic theory
of gases in order to understand the true nature of the flow.

Moreover, astrophysical flows are rife with multiphase interfaces and
shocks.  Shock interfaces are discontinuities in the fluid limit and
there are many accurate and elaborate schemes for their numerical
computation in the CFD pantheon.  However, in many cases of interest,
the dynamics of the particles in this interface regime is critical to
understanding the overall energetics through cooling and heating and
the observational signatures in the form of line and continuum
strength predictions and are unlikely to be well-described by the LTE
approximation.  Such calculations also require a kinetic theory
approach.

This paper describes an implementation of a Monte Carlo solution to
the collisional Boltzmann equation known as \emph{Direct Simulation
  Monte Carlo} (DSMC).  Algorithmically, it splits the full Boltzmann
equation into a purely collisionless left-hand side and a purely
collisional right-hand side and solves the two parts sequentially.
The solution of the former is solved is provided by a standard n-body
procedure.  The solution of the latter uses a space partition to
define interaction domains for the simulated gas particles that are
used to evaluate the Boltzmann collision integral (see
\S\ref{sec:algorithm}).  The implementation described here uses a
doubly nested octree for decomposing the spatial domain; the
first-level coarse-grained tree is used for load balancing and the
second-level fine-grained tree is used by each process to construct
interaction domains (see \S\ref{sec:parallel}).  This approach is much
faster than the brute-force approach of molecular dynamics but still
much slower than CFD.

Although DSMC will work even in the limit of dense gases
\citep{Alexander.etal:95}, we would like to use the full kinetic
approach only when needed owing to its computational expense.  A
number of such approaches have been proposed in the literature,
e.g. \citet{Garcia.etal:1999,Wissink.etal:2001a,Wijesinghe.etal:2004}
present adaptive mesh refinement schemes that use the Navier-Stokes
equations or DSMC depending on the regime.  The implementation in this
paper also takes a multiscale adaptive approach based on the local
density of particles: when the mean free path becomes small compared
the density scale, we use a particle-only approach
\citep{Pullin:80,Macrossan:2001} which solves the Navier-Stokes
equations in the limit of large particle number (see
\S\ref{sec:epsm}).  Furthermore, although a bit noisier and possibly
slower than the CFD--particle hybrid approaches, a particle-only
scheme is straightforwardly parallelized and mated with a traditional
n-body code.  Although we have thrown away the large-scale averaging
implicit in the continuum hydrodynamic equations by adopting a
particle-only approach, we have gained a method that is explicitly
stable (i.e. no Courant-Friedrichs-Lewy condition) and shock
boundaries are naturally resolved without need for artificial
viscosity or shock-capturing techniques.

We have seen in \S\ref{sec:tests} that this code correctly reproduces
the standard shock-tube tests and develops Kelvin-Helmholtz
instabilities that follow the analytic dispersion relation, both in
the continuum limit.  This approach does require an order of magnitude
(at least) more fluid particles in the continuum limit, although it
has the advantage of consistently transitioning to dilute gases,
correctly resolving shocks, and resolving phase boundaries.

The tests in Paper 2 use the classic fixed-composition cooling curves
in the LTE limit both to facilitate comparison with published results
and for the ease of implementation.  However, the traditional DSMC
implementation is based on individual particle cross sections.  It is
natural and straightforward to include multiple distinct species,
interactions, and excitations within the DSMC framework and
self-consistently compute the radiation spectrum from the gas in the
optically thin limit.  When the heating or cooling of the gas overall
depends on specific elemental or molecular lines from species with low
fractional number density (e.g. singly or multiply ionized oxygen,
carbon and nitrogen with $T>10^4$ K), these species may be included as
\emph{tracer} subspecies by solving rate equations in the QSS limit or
using weighting schemes with or without particle production.  As
described in \S\ref{sec:motivation} and \S\ref{sec:coolheat}, we are
currently testing a DSMC implementation that includes any specie whose
atomic data is included in the CHIANTI atomic data base
\citep{Dere.etal:1997,Landi.etal:2011}, including the standard plasma
cross sections.  Continuing to generalize the microphysics, it should
be possible to consistently include additional plasma physics by
adding the simultaneous solution of the electrostatic Poisson
equation.  Of course, the time-dependent solution of Maxwell's
equations is a very stiff challenging problem, but intermediate
charge-flow problems are tractable using DSMC.

Paper 2 applies the hybrid DSMC--n-body code from this paper to study
the effect of an ICM wind on a galaxy's ISM, commonly known as
\emph{ram pressure}.  As illustrated by Paper 2, DSMC may help us
understand the multiphase medium on small scales by enabling accurate
treatment of interfaces in the ISM.  Finally, on much larger scales,
DSMC can be used to simulate the dominant processes in intra-cluster
gas dynamics, such as the formation and interaction of bubbles,
conduction at interfaces, etc.

\section*{Acknowledgments}

This material is based upon work supported by the National Science
Foundation under Grant No. AST-0907951. 

\bibliographystyle{mn2e}

\begin{thebibliography}{43}
\expandafter\ifx\csname natexlab\endcsname\relax\def\natexlab#1{#1}\fi

\bibitem[{Agertz {et~al}\mbox{.}(2007)Agertz, Moore, Stadel, Potter, Miniati,
  Read, Mayer, Gawryszczak, Kravtsov, Nordlund, {et~al.}}]{Agertz.etal:07}
Agertz O. {et~al.}, 2007, Monthly Notices of the Royal Astronomical Society,
  380, 963

\bibitem[{Alexander, Garcia \& Alder(1995)Alexander, Garcia, \&
  Alder}]{Alexander.etal:95}
Alexander F.~J., Garcia A.~L., Alder B.~J., 1995, Phys. Rev. Lett., 74, 5212

\bibitem[{Begelman, Blandford \& Rees(1980)Begelman, Blandford, \&
  Rees}]{Begelman.etal:80}
Begelman M.~C., Blandford R.~D., Rees M.~J., 1980, Nature, 287, 301

\bibitem[{Bird(2007)}]{Bird:2007}
Bird G., 2007, in Notes from DSMC07 meeting, Santa Fe, September

\bibitem[{Bird {et~al}\mbox{.}(2009)Bird, Gallis, Torczynski, \&
  Rader}]{Bird.etal:09}
Bird G., Gallis M., Torczynski J., Rader D., 2009, Physics of Fluids, 21,
  017103

\bibitem[{Bird(1994)}]{Bird:94}
Bird G.~A., 1994, Molecular gas dynamics and the direct simulation of gas
  flows. Clarendon Press

\bibitem[{Birkinshaw(1991)}]{Birkinshaw:91}
Birkinshaw M., 1991, \mnras, 252, 73

\bibitem[{Black(1981)}]{Black:81}
Black J.~H., 1981, \mnras, 197, 553

\bibitem[{Bodenheimer {et~al}\mbox{.}(2006)Bodenheimer, Laughlin, Rozyczka, \&
  Yorke}]{Bodenheimer.etal:2006}
Bodenheimer P., Laughlin G.~P., Rozyczka M., Yorke H.~W., 2006, Numerical
  Methods in Astrophysics: An Introduction, Series in Astronomy and
  Astrophysics. Taylor \& Francis

\bibitem[{Boyd, Chen \& Candler(1995)Boyd, Chen, \& Candler}]{Boyd.etal:1995}
Boyd I.~D., Chen G., Candler G.~V., 1995, Physics of Fluids, 7, 210

\bibitem[{Cercignani(1990)}]{Cercignani:90}
Cercignani C., 1990, Mathematical Methods in Kinetic Theory, 2nd edn. Springer

\bibitem[{Cercignani(2000)}]{Cercignani:00}
Cercignani C., 2000, Rarefied gas dynamics. From basic concepts to actual
  calculations. Cambridge University Press

\bibitem[{Chandrasekhar(1961)}]{Chandrasekhar:1961}
Chandrasekhar S., 1961, Hydrodynamic and Hydromagnetic Stability. Oxford
  Univesity Press

\bibitem[{Chen \& Boyd(1996)}]{Chen.Boyd:1996}
Chen G., Boyd I.~D., 1996, Journal of Computational Physics, 126, 434

\bibitem[{Dere {et~al}\mbox{.}(1997)Dere, Landi, Mason, Monsignori~Fossi, \&
  Young}]{Dere.etal:1997}
Dere K., Landi E., Mason H., Monsignori~Fossi B., Young P., 1997, Astronomy and
  Astrophysics Supplement Series, 125, 149

\bibitem[{Fallavollita, Baganoff \& McDonald(1993)Fallavollita, Baganoff, \&
  McDonald}]{Fallavollita.etal:1993}
Fallavollita M., Baganoff D., McDonald J., 1993, Journal of Computational
  Physics, 109, 30

\bibitem[{Fritz(2001)}]{Fritz:2001}
Fritz J., 2001, Lectures in Mathematical Sciences, 18

\bibitem[{Gabriel {et~al}\mbox{.}(2004)Gabriel, Fagg, Bosilca, Angskun,
  Dongarra, Squyres, Sahay, Kambadur, Barrett, Lumsdaine,
  {et~al.}}]{Gabriel.etal:2004}
Gabriel E. {et~al.}, 2004, in Recent Advances in Parallel Virtual Machine and
  Message Passing Interface, Springer, pp. 97--104

\bibitem[{Garcia {et~al}\mbox{.}(1999)Garcia, Bell, Crutchfield, \&
  Alder}]{Garcia.etal:1999}
Garcia A.~L., Bell J.~B., Crutchfield W.~Y., Alder B.~J., 1999, Journal of
  computational Physics, 154, 134

\bibitem[{Garcia \& Wagner(2000)}]{Garcia.Wagner:2000}
Garcia A.~L., Wagner W., 2000, Journal of Statistical Physics, 101, 1065

\bibitem[{Gropp, Lusk \& Thakur(1999)Gropp, Lusk, \& Thakur}]{Gropp.etal:1999a}
Gropp W., Lusk E., Thakur R., 1999, Using MPI-2: Advanced features of the
  message passing interface, Vol.~2. MIT Press (MA)

\bibitem[{Hernquist \& Ostriker(1992)}]{Hernquist.Ostriker:92}
Hernquist L., Ostriker J.~P., 1992, ApJ, 386, 375

\bibitem[{Hopkins(2013)}]{Hopkins:13}
Hopkins P.~F., 2013, Monthly Notices of the Royal Astronomical Society, 428,
  2840

\bibitem[{Kendall \& Stuart(1983)}]{Kendall.Stuart:83}
Kendall M.~G., Stuart A., 1983, The advanced theory of statistics, 4th edn.,
  Vol.~1. C. Griffin

\bibitem[{Landi {et~al}\mbox{.}(2011)Landi, Del~Zanna, Young, Dere, \&
  Mason}]{Landi.etal:2011}
Landi E., Del~Zanna G., Young P., Dere K., Mason H., 2011, The Astrophysical
  Journal, 744, 99

\bibitem[{Macrossan(2001)}]{Macrossan:2001}
Macrossan M.~N., 2001, in AIP Conference Proceedings, Vol. 585, 22nd
  International Symposium on Rarefied Gas Dynamics, AIP, pp. 388--395

\bibitem[{Mo, van~den Bosch \& White(2010)Mo, van~den Bosch, \&
  White}]{Mo.etal:2010}
Mo H., van~den Bosch F., White S., 2010, Galaxy Formation and Evolution.
  Cambridge University Press

\bibitem[{Montanero \& Santos(1996)}]{Montanero.Santos:96}
Montanero J.~M., Santos A., 1996, Phys. Rev. E, 54, 438

\bibitem[{Navarro, Frenk \& White(1997)Navarro, Frenk, \&
  White}]{Navarro.Frenk.ea:97}
Navarro J.~F., Frenk C.~S., White S.~D.~M., 1997, \apj, 490, 493

\bibitem[{Park(1990)}]{Park:1990}
Park C., 1990, Non-Equilibrium Hypersonic Aerothermodynamics. A
  Wiley-Interscience Publication, pp. 255--268

\bibitem[{Perucho \& Lobanov(2007)}]{Perucho.Lobanov:07}
Perucho M., Lobanov A.~P., 2007, \aa, 469, 23

\bibitem[{Pullin(1980)}]{Pullin:80}
Pullin D.~I., 1980, J. Comput. Phys., 231

\bibitem[{Read, Hayfield \& Agertz(2010)Read, Hayfield, \&
  Agertz}]{Read.etal:10}
Read J., Hayfield T., Agertz O., 2010, Monthly Notices of the Royal
  Astronomical Society, 405, 1513

\bibitem[{Serikov, Kawamoto \& Nanbu(1999)Serikov, Kawamoto, \&
  Nanbu}]{Serikov.etal:1999}
Serikov V.~V., Kawamoto S., Nanbu K., 1999, Plasma Science, IEEE Transactions
  on, 27, 1389

\bibitem[{Sod(1978)}]{Sod:78}
Sod G.~A., 1978, J. Comput. Phys., 27, 1

\bibitem[{Sun {et~al}\mbox{.}(2011)Sun, Tang, He, \& Tao}]{Sun.etal:2011}
Sun Z.-X., Tang Z., He Y.-L., Tao W.-Q., 2011, Computers \& Fluids, 50, 1

\bibitem[{Toro(1999)}]{Toro:1999}
Toro E.~F., 1999, Riemann Solvers and Numerical Methods for Fluid Dynamics: A
  Practical Introduction, 2nd edn. Springer

\bibitem[{Wagner(1992)}]{Wagner:92}
Wagner W., 1992, J. Statist. Phys., 66, 1011

\bibitem[{Weinberg(1999)}]{Weinberg:99}
Weinberg M.~D., 1999, \aj, 117, 629

\bibitem[{Weinberg(2013)}]{Weinberg:13b}
Weinberg M.~D., 2013, \mnras, submitted

\bibitem[{Wijesinghe {et~al}\mbox{.}(2004)Wijesinghe, Hornung, Garcia, \&
  Hadjiconstantinou}]{Wijesinghe.etal:2004}
Wijesinghe H.~S., Hornung R.~D., Garcia A.~L., Hadjiconstantinou N.~G., 2004,
  J. Fluids Eng., 126, 768

\bibitem[{Wissink {et~al}\mbox{.}(2001)Wissink, Hornung, Kohn, Smith, \&
  Elliott}]{Wissink.etal:2001a}
Wissink A.~M., Hornung R.~D., Kohn S.~R., Smith S.~S., Elliott N., 2001, in
  Supercomputing, ACM/IEEE 2001 Conference, IEEE, pp. 22--22

\bibitem[{Zhang, Yao \& Li(2008)Zhang, Yao, \& Li}]{Zhang.etal:2008}
Zhang R., Yao W., Li J., 2008, Communications in Nonlinear Science and
  Numerical Simulation, 13, 2203

\end{thebibliography}

\label{lastpage}

\end{document}